# TRAINVERIFY: Equivalence-Based Verification for Distributed LLM Training

Yunchi Lu[†*], Youshan Miao[*], Cheng Tan[‡*], Peng Huang[†], Yi Zhu[*], Xian Zhang[*], Fan Yang[*]

[†]*University of Michigan*  [‡]*Northeastern University*  [*]*Microsoft Research*

**Abstract**

Training large language models (LLMs) at scale requires parallel execution across thousands of devices, incurring enormous computational costs. Yet, these costly distributed trainings are rarely verified, leaving them prone to silent errors and potentially wasting millions of GPU hours.

We introduce TRAINVERIFY, a system for *verifiable distributed training* of LLMs. Given a deep learning model's logical specification as the ground truth, TRAINVERIFY formally verifies that a distributed parallel execution plan is mathematically equivalent to it. Direct verification is notoriously difficult due to the sheer scale of LLMs which often involves billions of variables and highly intricate computation graphs. Therefore, TRAINVERIFY introduces shape-reduction techniques and a stage-wise parallel verification algorithm that significantly reduces complexity while preserving formal correctness. TRAINVERIFY scales to frontier LLMs, including the successful verification of the Llama3 405B and DeepSeek V3-671B training plans.

## 1 Introduction

Recent studies show that scaling deep neural network (DNN) generally yields better performance [42]. This triggers an "arms race" for large models. Training large models requires extensive resources, necessitating distributed training over a growing number of GPUs [25, 26, 36]. Latest large language models (LLM), for instance, are trained on clusters with thousands to tens of thousands of high-end GPUs for several weeks or months, costing millions of dollars [9, 32, 36].

Distributed training, however, is notoriously complex. It requires intricate spatial-temporal scheduling and coordination across a large set of devices using sophisticated parallelism techniques [44, 46, 59, 76]. The entire workflow is error-prone (§ 2.2). Worse still, the bug symptoms are often *silent*, such as causing wrong gradient updates, and improper scaling of the model states, which are difficult to detect and debug.

In this context, the cost of bugs becomes particularly high. Even a single bug can lead to wasting significant resources, financial loss, and low productivity for developers [18, 36]. Moreover, an incorrectly trained model can have far-reaching consequences, especially in critical applications where model accuracy is paramount [35, 41, 62].

Given the high stakes, there is an urgent need in verifying the correctness of distributed training. To achieve this goal, one option is to design a new training software stack that is correct-by-construction. While ideal, this is a daunting task from a verification perspective. Large-scale training involves a myriad of external dependencies as well as high concurrencies, both of which are challenging to verify [1, 10, 31, 73, 75]. The incompatibility with well-established training libraries [2, 22] and potential performance loss will also pose practical barriers.

In this work, we explore an alternative direction and propose a methodology that aims to provide strong correctness guarantees without rebuilding the whole training stack.

Our key insight is that the correctness of distributed training can be rigorously verified at the level of **execution plans**. By establishing what we call **parallelization equivalence**, we prove that the *execution plan* of a DNN model is equivalent to its *logical definition*. In other words, for *every* input, the distributed execution must produce the same output as the original model definition. This verifiability is achieved without running the training task at full scale.

Specifically, the *logical definition* of a DNN model can be represented as a data flow graph (DFG), where a node represents an operator like matrix computation, and an edge denotes a tensor produced by its upstream operators and consumed by its downstream operators. This representation is logical because operators and tensors are described mathematically. This logical form is usually carefully scrutinized through rigorous math reasoning and empirical validations, providing assurance of correctness [69]. When developers define a model in code, deep learning frameworks still internally use or provide ways to expose a DFG representation [65, 66].

The parallelization of a model decides how operators and tensors in the model are partitioned and scheduled across devices using parallelism techniques. This implementation logic is captured by a materialized *execution plan* [46], which can be structured as a parallelized DFG. Training systems use this plan to orchestrate the execution of a training task.

Parallelization equivalence ensures that for any input the parallelized DFG will produce the same output as the logical DFG does. Focusing on this property allows the verification effort to be manageable. It also supports existing frameworks, which makes the approach practical. At the same time, since the execution plan captures the essence of how a training task is parallelized, proving this property *eliminates* major classes of bugs that jeopardize the correctness of distributed training, providing strong guarantees(§ 8.3).

Several key research challenges arise: (1) *how to formulate the verification problem?* (2) *what representation to use to carry out the verification?* (3) *how to scale the verification to large DNNs that have hundreds of billions of parameters?*

To address these challenges, we first formally define the parallelization equivalence and its verification (§3). We then design TRAINVERIFY with multiple techniques to address needs of representation and scalability.

- *Symbolic DFG* (§5.1): TRAINVERIFY defines symbolic operators that encapsulate the mathematical foundations of modern deep learning operations. It converts a logical model and its execution plan into symbolic dataflow graphs and verifies the equivalence between these two representations.
- *Shape reduction* (§5.2) mitigates scale explosion in large models by reducing tensor shapes while preserving their structural and functional properties. It enables verification on smaller, manageable shapes and provably extends the results to larger shapes with the same structure.
- *Staged verification* (§5.3) addresses the intractability of verifying long symbolic expressions in DNNs by partitioning the DFGs into stages, each treated as a separate verification unit and verified concurrently. TRAINVERIFY then chains these units into an end-to-end proof by verifying input-output equivalence for dependent units and leveraging lineage metadata to ensure correctness across the entire network.

To our best knowledge, TRAINVERIFY is the first to offer *provably correct execution plans* for distributed training. It has been integrated into nnScaler [46], a state-of-the-art training framework. We identify the changes needed to make nnScaler amenable to verification. Specifically, we enhance the DFGs to incorporate all training-related computations. In addition, verification requires *lineage* to effectively map tensor values between the logical DFG and its parallelized counterpart. This information is discarded after parallelization in nnScaler, but is preserved in TRAINVERIFY, enabling equivalence checking.

Experiments show that TRAINVERIFY successfully verifies the executions plans for state-of-the-art large language models, LLaMA3 (8B/70B/405B) [36] and DeepSeek-V3 (16B/236B/671B) [34]. The verification finishes within 0.5 to a few hours for small and medium models, and up to 47 hours for the largest models. We also show that TRAINVERIFY can detect and eliminate major classes of real-world parallelization bugs in distributed training.

## 2 Background and Motivation

### 2.1 Distributed Training

At a high level, distributed training partitions the computational operations over high-dimensional data into multiple operations over smaller data, schedules the partitions among multiple devices *spatially*, executes them in a specific order *temporally*, and synchronizes state across different devices.

| Cause | MegatronLM | DeepSpeed | nnScaler |
|---|---|---|---|
| op-transformation | 16 | 18 | 19 |
| scheduling | 4 | 1 | 4 |
| communication | 8 | 13 | 5 |
| Total | 26 | 28 | 25 |

**Table 1.** *Parallelization-specific bugs* in distributed training systems, based on commit logs and GitHub issues (categories may overlap).

Different parallelization approaches have been proposed, including *data parallelism*, *tensor parallelism*, and *pipeline parallelism* [6, 38, 61]. Large-scale training in practice commonly employs a combination of these techniques, such as 3D parallelism [29, 36].

To parallelize a specific model, ML engineers typically adapt or handcraft detailed parallelization logic on top of a library (*e.g.*, Megatron-Core [10]) that provides APIs for common parallelism techniques. Recent training frameworks [46, 76] also automatically explore the search space to generate efficient parallelization strategies for a given model.

### 2.2 Error-Proneness of Parallelization

Despite its power, distributed training is inherently error-prone due to its significant complexities. It requires correctly partitioning the operators and tensors, scheduling them into the right devices, inserting proper communication operators, coordinating execution in the right order, and ensuring that all of this preserves the original model's semantics under *diverse* training configurations and parallelization schemes.

Indeed, as Table 1 shows, we study three state-of-the-art distributed training systems [46, 60, 61] and find that they all encounter bugs *specific to parallelization* in nearly every aspect of the workflow. For instance, in distributed data loaders, bugs can cause improper slicing and uneven distribution of global batches [5]. In forward and backward propagation, an operator can be incorrectly transformed or a tensor can be improperly partitioned [17]. Similarly, the calculation of metrics such as loss averaging and gradient norms can be affected if tensor shards are mistakenly treated as replicated or partitioned [4, 11, 16, 18]. Pipeline scheduling can also misorder operators from different microbatches or misallocate gradient updates [12, 14, 15]. In addition, a synchronization step may miss or use incorrect communication operations, *e.g.*, assigning wrong GPU ranks as sources or destinations, or using a wrong type of primitive [3, 13, 19, 20].

### 2.3 The Need for Rigorous Parallelization

What make parallelization errors particularly concerning is that they are often *silent*, so the training appears deceptively normal and the impact only becomes apparent later. For instance, incorrect gradient synchronization might only affect certain layers. These errors can evade detection until weeks into training, wasting valuable resources. Such errors are also challenging for developers to diagnose and debug. For example, a subtle incorrect loss scaling bug in MegatronLM took developers and users months of discussions to pinpoint [18].

It is also challenging for traditional testing to expose parallelization bugs for various inherent reasons. Deep learning relies on floating-point computation, which naturally introduces value drift during parallel training due to variations in the order of operations [63]. This drift is exacerbated by mixed-precision training, which employs low-precision data types and the use of diverse kernel implementations for the same operator due to efficiency. The numerical instability makes it challenging to distinguish normal numerical noises from actual errors, undermining methods like differential testing that directly comparing numerical results between unparallelized and parallelized training [50, 52]. For large-scale training that involves thousands of GPUs, it is also infeasible to obtain another full-scale single-device training result for comparison. Using a smaller-scale testing can easily miss bugs that only appear in full-scale training.

Given the importance of distributed training, the financial costs of training failures, the consequences of an incorrectly trained model, as well as the challenges in detection and diagnosis, it is imperative to provide formal correctness guarantees for parallelization. In other words, we should aim for *eliminating* parallelization bugs.

### 2.4 Insight: Verifying Parallelization Equivalence

While verifying the entire distributed training stack would be ideal, doing so is impractical. The stack includes numerous interacting components such as DNN compilers, kernel libraries, collective-communication runtimes, *etc.*, each of which is complex and often depends on opaque hardware vendor code [1, 10, 31, 73, 75]. The efforts required for verification would be prohibitively high. Compatibility and performance will also likely be sacrificed. Moreover, new parallelism techniques and operator optimizations emerge rapidly, making it difficult to maintain the verified system.

To alleviate this complexity, our key insight is that crucial correctness of distributed training resides in the **execution plan**—a representation of a full training iteration of the distributed model—generated by the parallelization framework. Once the plan is fixed, subsequent runtime merely executes the prescribed operations. Based on this insight, we shift the verification focus from the system to **parallelization equivalence (PE)**: the execution plan is equivalent to the model's logical definition. Both the execution plan and logical definition can be represented as data flow graphs (DFGs). PE requires that for all inputs, executing the parallelized DFG yields outputs that are identical to those produced by the logical DFG.

This verification approach offers several benefits. First, it can support existing training frameworks. It mainly requires the frameworks to provide DFG representation for the parallel execution plus lineage information. In our experience, this is not difficult. Second, the state space to verify becomes more tractable. We reason about one concrete plan each time, not the enormous space of all possible plans. Third, PE reasoning is symbolic over *real arithmetic*, making it immune to numerical noises. Last but not least, PE provides strong correctness guarantees. It covers all admissible inputs. An execution plan faithfully captures the computation and communication logic involved in a training iteration. Frameworks such as Alpa [76] and nnScaler [46] further materializes an execution plan into executable code. If PE holds, the distributed run is *functionally indistinguishable* from the carefully scrutinized logical model; if it fails, some bug must be present.

## 3 Parallelization Equivalence

In this section, we formulate and define the problem of verifying *parallelization equivalence*.

**Formulation.** Let $f : \mathcal{X} \to \mathcal{Y}$ denote a logical neural model, assuming for single-device execution. $\mathcal{X}$ and $\mathcal{Y}$ denote the universe of input and output tensors, respectively. A parallelization procedure $\mathcal{P}(\cdot)$ transforms $f$ into a model $g$ that executes on multiple devices. $g$ is parallelization equivalent to $f$ iff $\forall x \in \mathcal{X}, g(x) = f(x)$.

While the problem is theoretically broad and intractable—$f$ may involve arbitrary operators and $\mathcal{P}$ can be extremely complex—it is constrained in practice. We focus on $f$ as valid neural networks, *e.g.*, transformer-based models, and $\mathcal{P}$ as feasible parallelization plans.

Operations in $f$ include, but are not limited to, matrix operations (Linear, MatMul, Transpose, Reshape), activations (Softmax, SiLU), and others (Layernorm, Embedding). Parallelization $\mathcal{P}$ covers data parallelism [6], tensor parallelism [61], pipeline parallelism [38, 55], and their combinations. Expert [58], context [49], and sequence [45] parallelisms fall within our broader definition of tensor parallelism. $\mathcal{P}$ is applied to $f$ by inserting partitioning and communication operators (Chunk, AllReduce, AllToAll, etc).

**Graph Representation.** Both $f$ and $g$ can be encoded as data flow graphs (DFGs), where nodes represent operators and edges carry tensors. Then, the parallelization procedure $\mathcal{P}$ transforms the logical DFG denoted as $G_l := (V_l, E_l)$ into a parallelized DFG denoted as $G_p := (V_p, E_p)$, by partitioning or replicating operators and tensors, and adding aggregation and communication operators. $\mathcal{P}$ assigns and orders $G_p$ to multiple devices. Let $\mathcal{M} : E_l \to E_p^{\mathbb{N}}$ denote the mapping between each logical tensor and its partitioned counterparts.

**Definition 3.1** (Parallelization Equivalence). $G_p$ is parallelization equivalent to $G_l$ iff: $\forall t \in E_l, \forall x \in \mathcal{X}, t(x) = \bigoplus_{t_i \in \mathcal{M}(t)} t_i(x)$, where $\bigoplus$ is a composition operation determined by the parallelization scheme.

We refer to $G_l$ as **logical model**, and $G_p$ as the **execution plan** for the distributed training task. Both encompass all operations within a training iteration, including the forward pass, backward pass, optimizer steps, and metric computations.

```
1  def require_allreduce_dgrad(cfg) -> Bool: pass
2  def get_model_tp_group(cfg) -> ProcessGroup: pass
3  class LinearWithFrozenW(torch.autograd.Function):
4      def backward(self, gy):
5          gx = gy.matmul(self.w)
6          if require_allreduce_dgrad(self.cfg):
7              torch.distributed.all_reduce(gx,
8                  group=get_model_tp_group(self.cfg))
9          return gx
```

**Figure 1.** Code snippet of a backward computation.

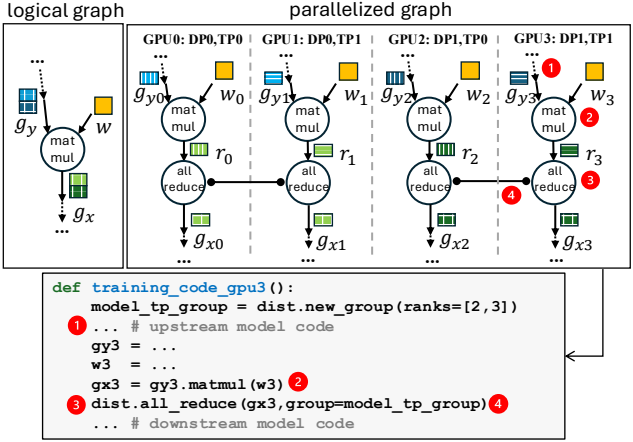

**Figure 2.** Logical and parallelized data flow graphs (simplified) for Figure 1 and the distributed training code for GPU 3.

**Example.** To illustrate how parallelization equivalence applies in practice, consider a custom autograd layer during back propagation, shown in Figure 1. It multiplies the upstream gradient $g_y$ with the fixed weight $w$ to obtain the gradient $g_x$. Figure 2 depicts the corresponding logical DFG.

Assume we train the same layer on four GPUs using 2-way data parallelism and 2-way tensor parallelism. Each GPU computes a partial result $r_i$ for the partitioned tensors. As Figure 2 shows, the parallelized DFG introduces an AllReduce operation to aggregate $r_i$ across tensor-parallel replicas.

Parallelization equivalence asserts that all the following algebraic equalities must hold for all valid inputs:
- $E_1 : g_y[\text{lower-half}] == g_{y2} + g_{y3}$
- $E_2 : g_x[\text{lower-half}] == r_2 + r_3$
- $E_3 : g_x[\text{lower-half}] == g_{x2} == g_{x3}$

These invariants are not just abstract properties; they capture classes of real-world bugs. To name a few:
- *Inconsistent tensor partitioning,* due to bugs that producer and consumer layers partition the same tensor inconsistently ❶. They will violate $E_1$.
- *Incorrect computation operator parallelization.* For example, if the operator is a custom operator, more complex than MatMul, involving rank-dependent computation, computation logic bugs can easily occur in ❷, violating $E_2$.
- *Missing or wrong communication operator.* For example, if require_allreduce_dgrad mistakenly returns False, the necessary AllReduce operation at ❸ will be missing.

Similarly, incorrect primitives may be introduced, *e.g.*, AllGather. All those cases will violate $E_3$.
- *Incorrect communication group.* For example, get_model _tp_group may be mistakenly assigned as a global group at ❹, *i.e.*, (0,1,2,3). They will also violate $E_3$.

## 4 Overview of TRAINVERIFY

To realize our verification methodology for DNN models in practice, we develop TRAINVERIFY, a system that provides provably correct execution plans for distributed training.

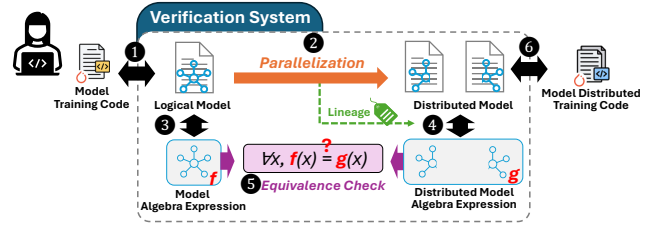

**Figure 3.** TRAINVERIFY System Overview.

As shown in Figure 3, developers write the code to describe the logical definition of a neural model. This code is usually device independent, or assumes running on a single virtual device with infinite power. Such code essentially represents a mathematical form of the model and can be rigorously scrutinized, reflecting the intended semantics of the model. To scale training to multiple GPUs, developers create a distributed training code, either manually implemented or leveraging auto-parallelization frameworks [46, 76].

TRAINVERIFY works in three steps. It takes both versions of training code as inputs and extracts two data flow graphs (DFGs) (❶, ❻). Nodes in the DFGs represent operations such as MatMul, AllReduce, etc., while edges carry data tensors. One (logical) DFG defines the logical model, while the other parallelized DFG encodes the parallelization procedure of how the model is partitioned and scheduled for distributed execution. TRAINVERIFY also completes each graph to capture a full training iteration.

Next, TRAINVERIFY symbolizes each graph by converting tensors into symbolic ones, and executable operations into symbolic computation (❸, ❹). This symbolic abstraction allows TRAINVERIFY to reason about the underlying algebraic semantics of the computation without being limited by specific data or affected by numeric noises. TRAINVERIFY uses lineage metadata (❷) to track how tensors and operations in the parallelized graph correspond to their counterparts in the logical model.

Finally, TRAINVERIFY checks the parallelization equivalence (PE) for the two symbolic DFGs by constructing formulas representing PE and using an SMT solver to formally verify that these formulas satisfy for all inputs (❺). If PE does not hold, TRAINVERIFY can output counter-examples.

**Scope.** To make the verification tractable, TRAINVERIFY has a focused scope. It treats the logical model as the specification and assumes it represents the desired semantics. It performs verification at the level of execution plans, which *algebraically* describes a complete training iteration. It does not target bugs that cause the training frameworks to crash before generating an execution plan, as such bugs can be handled by traditional tools. It focuses on verifying the parallel execution logic and eliminating correctness bugs essential to parallelization, such as incorrect transformation, wrong synchronization, and incorrect rank assignment. It does not aim to verify the runtimes for carrying out an execution plan, such as the kernel or communication libraries, or the operators' implementation details beyond their semantics (*e.g.*, internal buffer operations in optimizers and collective communication primitives).

## 5 System Design

In this section, we present the design of the TRAINVERIFY system, which enables the representation of Symbolic Data Flow Graphs (sDFG) for verification (§5.1). We also introduce two key techniques that enhance scalability: shape reduction (§5.2) and staged verification (§5.3).

### 5.1 Symbolic Data Flow Graph

As introduced in §2, deep learning models are commonly represented as data flow graphs in many DNN frameworks. However, these off-the-shelf graphs are insufficient for fully capturing the semantics of parallelized model trainings, due to two limitations: (1) they often omit critical computations involved in training, such as backpropagation, optimizer logic and metrics calculations, *e.g.*, gradient norm (gnorm); and (2) many operators in deep learning frameworks lacks formal mathematical definitions that are compatible with verification tools such as SMT solvers. Therefore, TRAINVERIFY extends conventional graphs into *symbolic data flow graphs* (sDFGs).

**Graph Completion.** A regular DFG usually represents the forward pass computation of a model. TRAINVERIFY reconstructs the backward pass by following automatic differentiation techniques [7, 28, 56] and incorporating customized implementations. TRAINVERIFY also replays optimizer and metric calculations in the graph, adhering to the implementation in distributed training systems [23, 46]. After completion, a DFG will include all the computations in a training iteration.

**Graph Symbolization.** TRAINVERIFY symbolizes the completed DFG into sDFG by replacing nodes and edges with algebraically defined symbolic tensors and operators. *Symbolic tensors* share the same shape with the original tensors in DFG, but with symbolic Real as elements without numeric instability, rather than typical FP32 or BF16 data types. *Symbolic operators* represent the same arithmetic operations over tensors as their DNN counterparts, including common PyTorch operators such as MatMul, AllReduce, etc. These operators are rewritten in TRAINVERIFY to be compatible with symbolic operands. For example, the symbolic MatMul is defined as: $C_{i,j} = \sum_k A_{i,k} \cdot B_{k,j}$ where $A$, $B$, and $C$ are symbolic tensors, and the computation is performed element-wise using symbolic expressions. This formulation allows the output tensor to be represented as an algebraic expression over the input tensors through symbolic operations. By recursively substituting input tensors with their corresponding symbolic expressions, one can compose consecutive operators—ultimately construct an algebraic expression representing the entire model.

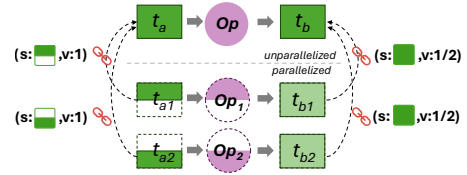

**Figure 4.** Lineage of tensors.

**Lineage.** *Lineage* is a data structure that preserves the semantics of parallelization, indicating how a tensor in the original logical model is partitioned or replicated during parallelization, analogous to DTensor [64] from PyTorch and many other similar concepts [46, 74, 76]. TRAINVERIFY also includes tensor lineage between the graph for the logical model definition and the corresponding distributed graph as tensors' property. As illustrated in Figure 4, operator Op is partitioned into sub-operators $Op_1$ and $Op_2$, with the input tensor $t_a$ spatially partitioned into sub-tensors $t_{a1}$ and $t_{a2}$. The lineage implies that $t_a ==$ concat($[t_{a1}, t_{a2}]$, dim = 0). Similarly, the output tensor $t_b$ is partitioned into $t_{b1}$ and $t_{b2}$, where the lineage maintains the same shape as $t_b$ but partial value, implying $t_{b1} + t_{b2} = t_b$.

Not all tensors have lineage, such as the tensors created during parallelization which lacks counterpart tensors in the graph for the logical model definition.

Lineage serves as a crucial bridge between the sDFG for logical model definition and the distributed sDFG during equivalence checking. Therefore, all input and output tensors must carry lineage information to enable end-to-end equivalence checking. The lineage of intermediate tensors will be used in stage-wise parallel equivalence verification (§5.3).

### 5.2 Shape Reduction

**Problem.** Modern DNN models contain tens or even hundreds of billions of variables, making verification computationally infeasible due to the sheer scale. The *shape reduction* mechanism reduces complexity while preserving verification fidelity.

**Insight.** Although DNN operators typically process large high-dimensional tensors, many operations exhibit redundancy. This redundancy arises from the SIMD (Single Instruction, Multiple Data) property inherent to DNN operators, which apply the same computation across different data elements of a tensor. As shown in Figure 5, elements $c_{1,1}$ and $c_{2,2}$ from output tensor are computed with the same algebra expression

## Algorithm 1 Inferring Minimum Reduced Shapes

**Input:** Topologically sorted parallelized graph $G_p$
**Output:** Tensors' reduced shapes shape_map

1: tensors ← $G_p$.data_inputs + $G_p$.weight_tensors
2: constraints ← ∅
3: **for all** $t \in$ tensors **do**
4:     **for** $d \leftarrow 1$ to $t$.ndim **do**
5:         $t$.rx_shape[$d$] ← init_symbolic_int()
6:         constraints.add(
7:             $1 \leq t$.rx_shape[$d$] $\leq t$.shape[$d$]) ❶
8: **for all** $op \in G$ **do**
9:     inputs ← $op$.inputs()
10:    outputs ← $op$.outputs()
11:    constraints.add(
            $op$.shape_align_cons(inputs, outputs)) ❷
12:    constraints.add(
            $op$.semantic_cons(inputs, outputs)) ❸
13: objective ← $\sum_{t \in \text{tensors}} \prod (t.\text{rx\_shape})$ ❹
14: model ← optimizer.minimize(objective)
15: shape_map ← {$t$ : model.eval($t$.rx_shape)
16:             for $t \in$ tensors}
17: **return** shape_map

but over distinct input sets:

$$c_{1,1} = a_{1,1} \cdot b_{1,1} + a_{1,2} \cdot b_{2,1} + a_{1,3} \cdot b_{3,1}$$
$$c_{2,2} = a_{2,1} \cdot b_{1,2} + a_{2,2} \cdot b_{2,2} + a_{2,3} \cdot b_{3,2}$$

**Figure 5.** DNN operator MatMul: different output elements $c_{1,1}$ and $c_{2,2}$ are calculated using the same function but on different inputs.

Therefore, verification may be performed on the same sDFG with only tensors' shape reduced, omitting redundant elements without compromising the verification outcome. We formally prove that verification on a reduced-shape sDFG is equivalent to verifying the original one. A proof sketch is provided in §6, with the full proof available in Appendix C.

**Minimum Shapes.** Shape reduction of a tensor must respect a series of constraints: (1) *Shape alignment*: For example, in a matrix multiplication of shape noatation MatMul([$M,K$] × [$K,N$]) = [$M,N$], the dimensions $M$, $K$, and $N$ occur across multiple tensors, so their corresponding sizes must match. It may also involve algebraic reasoning. For example, a reshape operation that converts a tensor from shape [$M,N$] to [$M,P,Q$] must satisfy the constraint $P \times Q = N$ to ensure the total number of elements remaining unchanged. (2) *Semantic intact*: reduced shapes must preserve the semantics of the corresponding operations. For example, in MatMul([$M,K$] × [$K,N$]) = [$M,N$], it should avoid reducing the dimension $K$ to 1 as invalidating the Add computation.

**Algorithm.** The procedure of shape reduction is shown in Algorithm 1. By collecting the minimum feasible shape for each tensor dimension, it computes model-wise minimum shapes based on these constraints.

The input $G_p$ is topologically sorted to ensure that, during graph traversal, each operator is visited only after its input tensors have been initialized or produced. Then, determining model tensors' minimum shapes is formulated as an integer quadratic problem subject to these collected *shape constraints*: ❶ Each dimension's minimum size should between 1 and the original size; ❷ Operator enforces *shape alignment* constraints over input and output tensors' dimensions. ❸ Operator enforces *semantic intact* constraints over tensor dimensions; The optimization goal ❹ uses the total tensor volume to approximate solver complexity, while allowing alternative objectives for solver efficiency. After determine minimum shapes of $G_p$, the tensor shapes in the original, unparallelized graph are recovered using *lineage*.

### 5.3 Staged Verification

**Problem.** While shape reduction mitigates complexity of large tensors, modern LLMs still face an additional challenges of their extreme depth. Verifying equivalence of such models experience exponential complexity growth due to their combinatorial nature. As model depth increases, symbolic expressions become intractable, often causing solvers to return unknown or fail to terminate due to memory overflow, even for comparing a single element in tensors. This highlights the need for an effective approach—decomposing the end-to-end verification into smaller, tractable sub-problems.

**Insight.** Inspired by pipeline parallelism, where a model is divided into multiple stages, the equivalence verification of the entire model can be similarly decomposed into verifying each of its stages individually. If, for every stage in the logical (non-parallel) model, we can verify that it equals to its corresponding stage in the parallelized model, then it naturally follows that the entire model, formed by composing these stages sequentially, is also equivalent. Given that the available *lineage* specifies the parallelization semantics of input and output tensors for each stage, their equivalence can be verified independently. Formally, if two corresponding stages $f$ and $f'$ are equivalent for all inputs, and similarly for $g$ and $g'$, then their composition must also be equivalent:

$$\forall x_1, f(x_1) = f'(x_1) \land \forall x_2, g(x_2) = g'(x_2)$$
$$\Rightarrow \forall x, f(g(x)) = f'(g'(x))$$

**Design.** TRAINVERIFY introduces *staged verification*. Logical graph $G_s$ and multi-device graph $G_p$ are aligned, and partitioned into *stages*. Each stage $S$ consists of two corresponding subgraphs, $S.G_s$ and $S.G_p$, drawn from the two respective graphs. If all stages pass the verification, the process guarantees end-to-end equivalence between the two graphs.

**Algorithm 2** Stage Parallel Verification

**Input:** Logical graph $G_s$, multi-device graph $G_p$, lineage $L$, reduced shapes $X$
**Output:** A Boolean indicating whether $G_s$ is equivalent to $G_p$.

1: $R \leftarrow$ equivalence of $G_s$ and $G_p$ inputs  ▷ global relation pool
2: stages $\leftarrow$ align_and_partition($G_s, G_p, L$)
3: **for** stage $S$ in stages **do**
4:     in_eq $\leftarrow S.G_s$.inputs $\stackrel{L,X}{==} S.G_p$.inputs
5:     out_eq $\leftarrow S.G_s$.outputs $\stackrel{L,X}{==} S.G_p$.outputs
6:     **async run** WORKER($S$, $R$.snapshot(), in_eq, out_eq)
7:     $R$.add(out_eq)
8: **return** all([$S$.worker.result()==**True** for $S$ in stages])
9:
10: **function** WORKER($S$, $\mathcal{R}$, eq$_{in}$, eq$_{out}$)
11:     $S$.init_symbolic_variables(shape=$X$)
12:     **if** (eq$_{in}$ is **not** AlwaysTrue **given** $\mathcal{R}$) **then**
13:        **return False**
14:     outputs $\leftarrow$ symbolically execute $S.G_s$ and $S.G_p$
15:     **if** (eq$_{out}$ is **not** AlwaysTrue **given** outputs $\cup$ eq$_{in}$ ) **then**
16:        **return False**
17:     **return True**

TRAINVERIFY obtains *tensor alignment* via lineage analysis, identifying bundles that each contains an original tensor in $G_s$ along with its counterparts in $G_p$. In determining this alignment, TRAINVERIFY traverses $G_s$ in topological order while synchronously traversing $G_p$.

TRAINVERIFY then *determines stages* by repeatedly applying backward slicing [72] on the dual graphs. Tensor alignment aids in aligning operators and guides the partitioning of the entire graph into stages. Specifically, during a topological traversal of $G_s$, whenever an unvisited tensor with lineage information is encountered, it is designated as the output tensor of a new stage $S$. A backward trace is then initiated from this tensor to identify the corresponding input tensors of the stage in $G_s$. This trace follows the reverse direction of graph edges and terminates upon reaching tensors that also carry lineage, thereby defining the boundaries of a subgraph in $G_s$. Using the lineage of each boundary tensor, the corresponding boundaries in $G_p$ are also identified. The stage $S$ is then constructed by extracting the associated subgraphs from both $G_s$ and $G_p$. Subsequent stages are determined by continuing the traversal over the remaining unvisited portions of $G_s$ and repeating this process. The procedure eventually terminates at the model's output tensors, yielding a complete partitioning of all stages for verification.

**Parallel Solving.** Algorithm 2 shows an overview of stage-parallel verification. The global relation pool $R$ manages verified relationships among tensors (line 1). The input graphs are partitioned into stages following aforementioned process (line 2). For each stage, the desired equivalence of both input and output is pre-encoded by lineage (line 4-5), then workers can be issued sequentially and run in parallel (line 3,6). After a stage worker is issued, its target output equivalence is assumed correct and added to the global relation pool $R$ (line 7). Subsequent workers are issued without waiting for prior ones to complete. TRAINVERIFY uses a process pool for worker management, with a configurable concurrency limit. All workers are later synchronized at a barrier to ensure successful completion (line 8).

In each stage's parallel worker, symbolic variables for inputs are instantiated based on precomputed reduced shapes $X$ (line 11). Input equivalence is checked against previously verified relations (line 12). If the equivalence condition does not always hold, then algorithm terminates and returns False (line 13). Output equivalence is verified based on in-stage computation after applying the corresponding operators (line 14-16). The verification goals, including losses, final gradient updates, and metrics, correspond to ensuring the desired output equivalence at their affiliated stages.

**Supporting Customized Equivalence Approximations.** Parallelization of certain model architectures may intentionally introduce approximate mathematical equivalence to optimize performance. For instance, in distributed training of ResNet models, many implementations adopt per-device local Batch-Norm as an approximation of global BatchNorm, omitting the additional cross-device averaging that is implicitly embodied in the formulation for the logical model definition. Similarly, in distributed training of MoE LLMs such as DeepSeek-V3, routing is often restricted to a subset of nodes rather than spanning the entire cluster in order to reduce communication costs. TRAINVERIFY supports such approximations in staged verification by overriding the approximate computation (line 14) with its strict-equivalence version, when configured as allowable by the user. For approximated BatchNorm, the overridden computation restores the omitted cross-device averaging. For MoE routing, it bypasses the logic that restricts routing to a subset of nodes.

## 6 Proof sketch

This section presents a proof sketch for the correctness of the shape reduction method—*verified parallelization equivalence for shape-reduced plans extends to their full-size counterparts*. The complete proof is in Appendix C.

DNN operators are predominantly SIMD (Single-Instruction Multiple-Data), performing repeated, homogeneous computations (a *kernel function*) over array elements. This SIMD property is central to enabling our shape reduction. We begin by introducing formal definition of the SIMD function.

**Definition 6.1** (Kernel function). Let $f(\mathbf{x}) \rightarrow \mathbf{y}$ be a function. A kernel function $\theta$ associated with $f$ takes a subtensor from $\mathbf{x}$ and outputs a scalar value. Formally,

$$\theta_f : \mathbb{R}^k \rightarrow \mathbb{R},$$

where $k$ is the size of the subtensor of $\mathbf{x}$.

**Definition 6.2** (Dependency mapping). Let $f(\mathbf{x}) \rightarrow \mathbf{y}$ be a function. A dependency mapping $\tau$ associated with $f$ is a function that maps each index $i$ in $\mathbf{y}$ to a list of indices in $\mathbf{x}$.

$$\tau_f : idx(\mathbf{y}) \to [idx(\mathbf{x})],$$

where $idx(\cdot)$ is the indexing function of the tensor.

**Definition 6.3** (SIMD function). A function $f(\mathbf{x}) \to \mathbf{y}$ is a SIMD function if, each output element $\mathbf{y}[i]$ is computed as:

$$\mathbf{y}[i] = \theta(\mathbf{x}_1, \mathbf{x}_2, \ldots, \mathbf{x}_k),$$

where $\theta$ is the kernel function of $f$, and

$$\mathbf{x}_j = \mathbf{x}[\tau(i)[j]], \quad 1 \le j \le k$$

where $\tau$ is the dependency mapping of $f$.

A SIMD function $f$ is characterized by its dependency mapping $\theta_f$ and kernel function $\tau_f$. In compact notation, we write its operation as $\mathbf{y}[i] = \theta_f(\mathbf{x}[\tau_f(i)])$.[I]

We have two key observations for LLM operators.

***Observation 1: LLM operators are SIMD functions.*** We observe that all layers in the transformer architecture are SIMD, including Feed Forward layers, Multi-Head Attention layers without masking (masks discussed in Appendix A), Add & Norm layers, ReLU, Softmax, and Residual addition.

As an example, matrix multiplication (i.e., MatMul) can be expressed as an SIMD function. Given two matrices $A \in \mathbb{R}^{m \times p}$ and $B \in \mathbb{R}^{p \times n}$, the resulting matrix $C \in \mathbb{R}^{m \times n}$ has elements $c_{i,j}$ (short for $C[i][j]$) computed by: $c_{i,j} = \sum_{k=1}^{p} a_{i,k} \cdot b_{k,j}$. MatMul is a SIMD function because it has

- a kernel function:

$$\theta(a_{i,1}, \ldots, a_{i,p}, b_{1,j}, \ldots, b_{p,j}) = \sum_{k=1}^{p} a_{i,k} \cdot b_{k,j};$$

- a dependency mapping for each input tensor:

$$\tau^A(i,j) = [(i,k) | 1 \le k \le p],$$
$$\tau^B(i,j) = [(k,j) | 1 \le k \le p],$$

where $\tau^A$ and $\tau^B$ are dependency mappings for input matrix A and B;

- and MatMul can be expressed as:

$$c_{i,j} = \theta(A[\tau^A(i,j)] \oplus B[\tau^B(i,j)]),$$

where $\oplus$ represents list concatenation.

***Observation 2: LLM operators have dependency mappings that can be expressed as linear combinations.*** This property is intuitive, as the "striding" of kernel functions across tensors typically occurs at regular, constant intervals [51]. Consequently, when the input to the dependency mapping—corresponding to the output tensor's index—changes, the resulting input indices change linearly. That is, the mapping takes the linear form: $\tau(i) = M \cdot i + b$.

---

[I] For brevity, we assume a single-tensor input for function $f$ throughout this paper. But our method can cover multiple input and output tensors.

For example, in the above MatMul case, the dependency mapping $\tau^A$ can be written as a linear form:

$$\tau^A(i,j) = M_A \begin{bmatrix} i \\ j \end{bmatrix} + b_A, \quad M_A = \begin{pmatrix} 1 & 0 \\ 1 & 0 \\ \vdots & \vdots \\ 1 & 0 \end{pmatrix}_{p \times 2}, \quad b_A = \begin{pmatrix} 0 & 1 \\ 0 & 2 \\ \vdots & \vdots \\ 0 & k \end{pmatrix}_{p \times 2}$$

**Proof sketch.** We establish the correctness of TRAINVERIFY's shape reduction by proving the equivalence between two data flow graphs (DFGs) on a subset of representative dimensions implies equivalence across all dimensions. We denote the logical and transformed DFGs (before and after applying parallelization) as functions $f$ and $g$, respectively. Before presenting the main theorem, we begin with two equivalence definitions that serve as the foundation for the proof.

**Definition 6.4** (Mapping permutation equivalence). For two dependency mappings $\tau_1$ and $\tau_2$, we call them mapping permutation equivalence, denoted $\tau_1 \cong_P \tau_2$, if there exists a permutation function $P$, such that

$$\forall i, \quad \tau_1(i) = P(\tau_2(i))$$

Mapping permutation equivalence captures LLM operators with commutative properties, where permuting the inputs does not affect the output. Similarly, we need to define a corresponding equivalence relation for kernel functions.

**Definition 6.5** (Kernel permutation-set equivalence). For two kernel functions $\theta_1$ and $\theta_2$, we call them kernel permutation-set equivalence, denoted $\theta_1 \cong_Q \theta_2$, if there exists a non-empty set $Q$ of permutation functions, such that

$$\forall P \in Q, \forall \mathbf{x}, \quad \theta_1(\mathbf{x}) = \theta_2(P(\mathbf{x}))$$

**Lemma 6.6.** *For SIMD functions $f$ and $g$,*

$$\theta_f \cong_Q \theta_g \land \tau_f \cong_P \tau_g \land P \in Q \implies f = g$$

*Proof.*

$$\begin{aligned}
\forall \mathbf{x}, \forall i, f(\mathbf{x})[i] &= \theta_f(\mathbf{x}(\tau_f(i))) && \text{[by Definition 6.3]} \\
&= \theta_g(P(\mathbf{x}[\tau_f(i)])) && \text{[by } \theta_f \cong_Q \theta_g \land P \in Q\text{]} \\
&= \theta_g(\mathbf{x}[P(\tau_f(i))]) && \text{[by tensor indexing rules]} \\
&= \theta_g(\mathbf{x}[\tau_g(i)]) && \text{[by } \tau_f \cong_P \tau_g\text{]} \\
&= g(\mathbf{x})[i]
\end{aligned}$$

Because for any input $\mathbf{x}$, $f(\mathbf{x})$ and $g(\mathbf{x})$ produce the same result, therefore $f = g$. □

Next, we prove that dimension reduction also applies to reductional operations such as sum. A reductional function $f : \mathbb{R}^n \to \mathbb{R}$ returns a single output element from processing a reduction operation among all elements in the input tensor, with the reduction operation satisfying the commutative and associative laws.

**Definition 6.7** (Reductional function). For an input tensor $X \in \mathbb{R}^n$, the reductional function $f$ applies a binary operation $\odot$ to all elements of $X$ such that:

$$f(X) = x_1 \odot x_2 \odot \cdots \odot x_n,$$

and $\odot$ satisfies commutativity ($a \odot b = b \odot a$) and associativity (($a \odot b) \odot c = a \odot (b \odot c)$).

> Theorems 6.8 and 6.9 prove the *essence* of shape reduction: verified parallelization equivalence on shape-reduced models *faithfully extends* to the original models.

**Theorem 6.8.** *Given reductional functions $f$ and $g$,*

$\forall \mathbf{x} \in \mathbb{R}^2, f(\mathbf{x}) = g(\mathbf{x}) \implies \forall \mathbf{x} \in \mathbb{R}^n, n \geq 2, f(\mathbf{x}) = g(\mathbf{x}).$

**Theorem 6.9.** *Let $f$ and $g$ be functions composed by LLM operators.*

$\exists i, \forall \mathbf{x}, f(\mathbf{x})[i] = g(\mathbf{x})[i] \implies \forall i, \mathbf{x}, f(\mathbf{x})[i] = g(\mathbf{x})[i]$

Theorem 6.8 can be easily proved by mathematical induction. Below shows a proof sketch for Theorem 6.9.

1. Our solver proves the precondition: there exists some dimension of the output tensor (namely, $i$) that both $f$ and $g$ produce the same result for any input $\mathbf{x}$.
2. By the precondition, we can derive that there exists a non-empty set $Q$ such that $\theta_f \cong_Q \theta_g$. We prove this by contradiction—if $\theta_f$ and $\theta_g$ are not kernel permutation-set equivalence, then there must exist some input $\mathbf{x}'$ where $\theta_f(\mathbf{x}') \neq \theta_g(\mathbf{x}')$, which contradicts the precondition.
3. By our observation 2 and the precondition, we can prove $\tau_f \cong_P \tau_g$. From the precondition, we can prove $\exists P, \tau_f(i) \cong_P \tau_g(i)$. By observation 2, we prove that the $P$ applies to all dimensions (i.e., $\tau_f \cong_P \tau_g$) due to the linear algebra.
4. Then, using the precondition and $\tau_f \cong_P \tau_g$, we prove $P \in Q$ by contradiction: assume $P \notin Q$, then $\exists \mathbf{x}', \theta_f(\mathbf{x}') \neq \theta_g(P(\mathbf{x}'))$, which contradicts the precondition.
5. Finally, by Lemma C.9, we prove $f = g$ because $\theta_f \cong_Q \theta_g$ (proved in step 2) and $\tau_f \cong_P \tau_g$ (proved in step 3) and $P \in Q$ (proved in step 4).

## 7 Implementation

We implement TRAINVERIFY in Python with 6,000 lines of code, using z3-solver [33] for equivalence check.

**Algebraic Expression of Models.** TRAINVERIFY leverages nnScaler, a state-of-the-art distributed training framework [46] that adopts graph-based parallelization. It traces model code written for single-GPU training to obtain an intermediate representation, then compiles it into a parallelized graph (IRGraph), and finally emits the corresponding distributed training code.

*Building full-fledged execution plans.* TRAINVERIFY naturally derives the multi-device plan from the IRGraph, without tracing the distributed code. The IRGraph encodes both forward and backward passes, with nodes representing PyTorch operators [57], communication primitives, and custom operators. As the distributed gnorm computation resides in static code outside the graph, TRAINVERIFY completes the execution plan by injecting semantically equivalent logic as custom operators. The current version of TRAINVERIFY supports only ZeRO Stage 1 [59], and thus abstracts the optimizer as a local gradient update operation. The logical model is obtained by invoking nnScaler to emit a single-GPU execution plan, whose correctness is guaranteed by nnScaler and complemented similarly. Parallelization equivalence is then verified between the single-device and multi-device execution plans.

*Tracking tensor lineage between dual graphs.* TRAINVERIFY infers tensor lineage from nnScaler's metadata, where the index mapping specifies how a sub-tensor is sliced from the full tensor, and the value mapping indicates whether aggregation with sibling tensors is required for reconstruction. However, this metadata presents two limitations. First, corresponding tensors across the dual graphs may have mismatched IDs. TRAINVERIFY resolves this by using source code as beacons to match operators and align their associated tensors. Second, both index and value mappings are scoped to a single microbatch within one data pipeline, which leads to inconsistent parallelization semantics between the graphs. To resolve this, TRAINVERIFY shifts index mappings based on DP rank and microbatch ID, to reflect the global batch context; and uses SSA-based tensor IDs along with static analysis of communication nodes to identify value-wise aggregation across data pipelines.

*Adapting operators for shape-reduced symbolic tensors.* Operators in the IRGraphs are not directly applicable to symbolic variables. TRAINVERIFY implements shape-reduction rules and symbolic adaptations for the operators used in GPT [30], Llama 3, and DeepSeek-V3 models (Table 4 in Appendix A lists them). TRAINVERIFY rewrites both forward and backward computations. The rewritten operation logic strictly adheres to their original mathematical definitions.

*Extension for new model architectures.* Adaptation is only needed if new operators are introduced. The rewriting effort is moderate, primarily involving the use of NumPy [21] and Python's built-in operators to emulate PyTorch and communication operations. This convenience is enabled by Z3 variables supporting Python operator overloading, allowing symbolic expressions to behave like native types (e.g., int, float).

Appendix A describes how we address additional practical challenges, without compromising correctness, in shape reduction and adapting non-SIMD operators (*e.g.*, the embedding layer and topK operator).

**Adaptation to Other Frameworks.** TRAINVERIFY defines a *graph interface* and a *solver interface* to decouple from both nnScaler and Z3, as shown in Figure 6. To enable verification, a parallelization framework must provide an SSA computation graph enriched with tensor lineage. This may require framework-specific knowledge when such representations are not natively supported. The modular design of stage checker also facilitates switching between symbolic engines—*e.g.*, SymPy [53] or Gurobi [8]—to support operator rewriting and solver-specific optimizations.

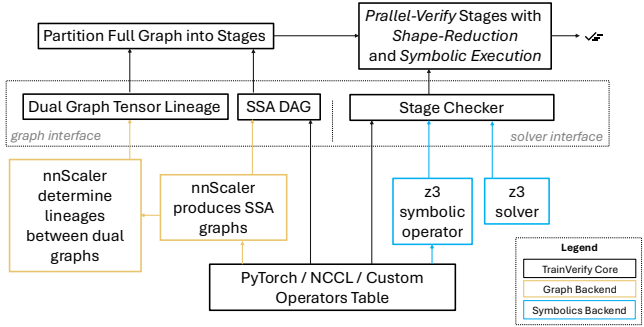

**Figure 6.** Implementation Overview.

**Accelerating Solving.** To enable practical verification time and memory usage for extremely large execution plans, we design several solver-specific optimizations.

*Efficient Equivalence Checking.* We design a hybrid equivalence checker. It first applies a superficial but efficient expression comparison: if the expressions are equal under simplification, the checker short-circuits without invoking the solver; otherwise, it falls back to the SMT solver. This design exploits the fact that expression comparison is lightweight and significantly faster than formal solving without false positives, though it may yield false alarms due to its superficial nature. To preserve soundness, any case that fails the expression checker is backed by the SMT solver. TRAINVERIFY implements expression checker using `z3.Tactic("solve-eqs")` [24], which simplifies equations via variable elimination and often reduces the problem to an empty conflict core when expressions are equivalent. The SMT solver used is the Z3 solver.

*Lazy Symbolization.* TRAINVERIFY instantiates Z3 symbolic variables only within parallel stage solvers, while maintaining only metadata in the main process. This design addresses two issues. First, symbolic variables and expressions are costly under Python `fork`, even with copy-on-write, and can add several seconds to each `fork` operation. Second, Z3 requires variables and solvers to be initialized within context tables, which have limited capacity (`unsigned_int`), making it impractical to store all symbolic tensors globally. Managing metadata alone also significantly reduces memory overhead.

## 8 Evaluation

We answer the following questions: (1) Can TRAINVERIFY verify large parallelized graphs of real-world distributed settings? (§8.1); (2) How does TRAINVERIFY scale with training parameters? (§8.2); (3) What classes of bugs can TRAINVERIFY eliminate? (§8.3)

### 8.1 Verifying Real-World LLM Parallelization

To demonstrate TRAINVERIFY's practicality, we experiment on verifying execution plans for LLaMA3 and DeepSeek-V3 models under various real-world setup.

**Setup.** The evaluated models' dataflow graphs are generated using nnScaler and verified by TRAINVERIFY on machines equipped with 4 NVIDIA A6000 GPUs, a 48-core Intel(R) Xeon(R) Silver 4310 CPU, 1 TB memory, and 2TB swap.

An overview of the parallelization schemes is presented in Table 2, and detailed model specifications are provided in Appendix B. We evaluate the LLaMA3 models at 8B, 70B, and 405B scales, with plans configured for up to 8192 GPUs, which follows the production setup [48, 68]. We evaluate the DeepSeek-V3 models at 16B, 236B, and 671B scales, with plans configured for up to 2048 GPUs. Expert parallelism is treated as a form of tensor parallelism in nnScaler, and thus shares the same degree as TP. As nnScaler currently enforces orthogonality between parallelism strategies, the experiment D3 employs a same-scale but different parallelization to approximate the official setup [34].

**Results.** Table 3 shows the verification time. For moderate-scale plans, such as L1 and D1, TRAINVERIFY completes verification within half an hour. For large-scale plans, such as L3 and D3, verification takes up to two days. The increased cost arises from both the enlargement of execution plans and the reduced solver parallelism due to memory constraints, which are further discussed in §8.2. While this cost is non-trivial, it remains acceptable given the correctness guarantees provided, especially when compared to the extended training durations, often several weeks, required for models at this scale.

### 8.2 Scalability

We evaluate how TRAINVERIFY verification time scales under various parallelization and model configurations. The base model used for variable-controlled experiments is LLaMA3-8B. Each experimental group is averaged over 5 independent runs. Detailed configurations are provided in Appendix B.

We observe Z3 exhibits variance in solving time depending on its random seed. Figure 7a shows TRAINVERIFY's performance with different Z3 random seeds, with variance observed only in the shape reduction solving time. The data point with seed 0 corresponds to the 512 global batch size point in Figure 7b (both are marked with ■). For all other experiments in this section, we use the default seed 0 without fine-tuning.

**Invariance to Original Tensor Shapes.** As shown in figs. 7b to 7e, the verification time of TRAINVERIFY, except for variance introduced by shape reduction solving, is independent of the actual tensor sizes. This invariance is achieved through the shape reduction technique, which effectively compresses the shapes of symbolic tensors in TRAINVERIFY's graphs. The compressed shapes depend solely on partitions rather than their original dimensions, the verification time remains constant regardless of variations in these parameters. For attention heads, increasing the number logically introduces more partitions. However, in efficient implementations, each GPU stacks its local heads and computes them with batched `matmul`. Thus, a larger number of attention heads is treated as reducible and does not impact the verification cost.

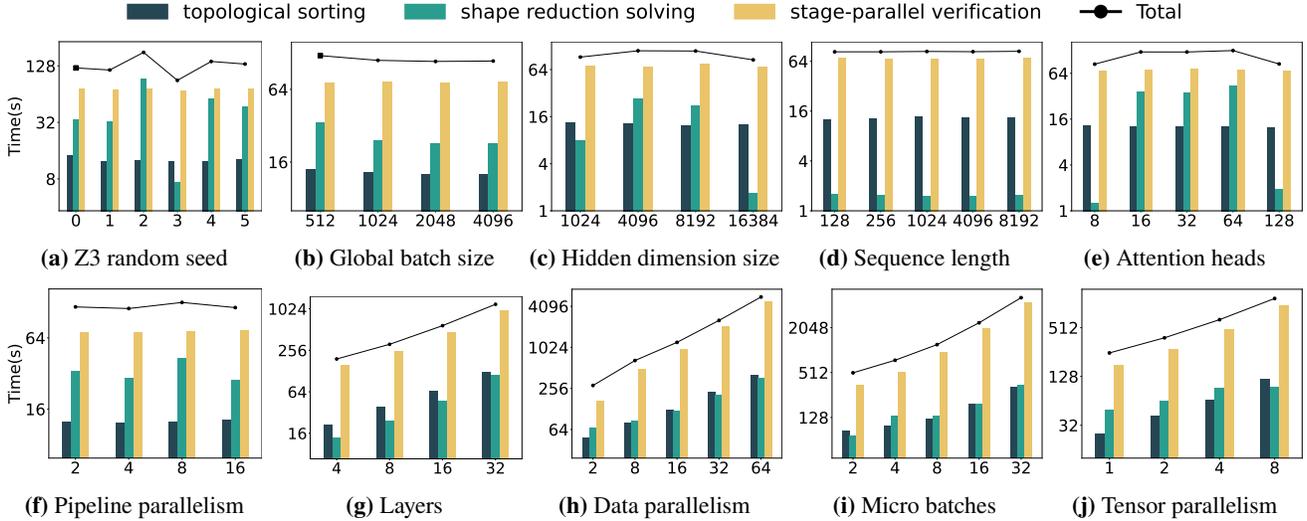

**Figure 7.** TRAINVERIFY's performance trends regarding different training configurations. The y-axes use a $\log_2$ scale. Bars indicate the time breakdown by component, while lines represent the end-to-end verification time.

| Exp. ID | Model | Layers | DP | TP | PP | NM |
|---|---|---|---|---|---|---|
| L1 | Llama3-8B | 32 | 512 | 1 | 1 | 1 |
| L2 | Llama3-70B | 80 | 16 | 8 | 4 | 32 |
| L3 | Llama3-405B | 126 | 64 | 8 | 16 | 16 |
| D1 | DS-V3-16B | 27 | 16 | 4 | 2 | 16 |
| D2 | DS-V3-236B | 60 | 16 | 8 | 4 | 16 |
| D3 | DS-V3-671B | 61 | 32 | 8 | 8 | 16 |

**Table 2.** Evaluated real-world large models.

|  | L1 | L2 | L3 | D1 | D2 | D3 |
|---|---|---|---|---|---|---|
| Solver Parallelism | 30 | 30 | 4 | 30 | 16 | 8 |
| End-to-end Time | 0.5h | 7.5h | 47h | 0.5h | 3.5h | 31h |

**Table 3.** Verification time for the evaluated models.

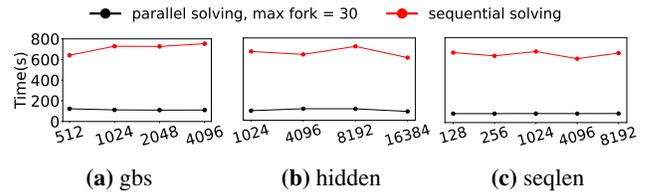

**Figure 8.** Verification time with vs. without stage parallelism.

**Linear Complexity to Parallelization.** Stage-parallel verification time consists of two main components: (1) the *main process time* for iterating over stages and inspecting tensor shard metadata to prepare inputs for the solver; (2) the *parallel solver time* for initializing symbolic variables, computing, and checking equivalence of each stage in the workers.

When resources are not constrained, the solvers are sufficiently parallelized, making the main process time dominate the verification time. Pipeline parallelism introduces a sub-linear increase in verification time, as Figure 7f shows, since additional stages contribute only a small number of communication operators. Increasing model layers adds operators sequentially, leading to a linear increase in the number of stages and time, as shown in Figure 7g. For data parallelism, microbatching, and tensor parallelism, the verification time grows linearly due to increased tensor copies proportional to the parallelism degree, along with the introduction of a small number of communication nodes, as figs. 7h to 7j show.

As the degree of parallelism increases, parallel solver time gradually dominates the verification time. This is because higher parallelism can cause the number of symbolic variables to grow quadratically, leading to quadratic or even greater complexity in solver. For instance, in experiment L3 (LLaMA3–405B in §8.1), the average main process time is 4.3s per stage, whereas the average solver time reaches 374s per stage, causing the main process to stall while waiting for worker completion. Similar trends are observed in experiments D2 and D3. While the dominance of solver time introduces deviations from linear scaling, the overall verification time remains upper-bounded by the statistics reported in §8.1.

**Stage-Parallel Verification.** We demonstrate the benefit of stage-parallel design by comparing it against a sequential configuration where solver parallelism is disabled. Figure 8 shows the results. Although these experiments use a relatively small-scale setup (LLaMA3 8B with 8 GPUs) where individual solvers are already lightweight, disabling solver parallelism increases the end-to-end verification time from under 2 minutes to over 10 minutes.

## 8.3 Eliminating Broad Categories of Bugs

To understand whether TRAINVERIFY addresses correctness issues that arise in real training systems, we consider broad categories of bugs that have been found in popular parallelization frameworks [46, 60, 61] and discuss whether enforcing parallelization equivalence eliminates them.

1. **Incorrect communication operators.** Such as missing `AllReduce` operations required for gradient synchronization, or activating unnecessary communication primitives.
2. **Incorrect device assignment.** Such as miscalculating the participating GPUs when initializing communication groups; or confusing the per-data-parallel mesh with the global device mesh.
3. **Incorrect partitioning of computational operators.** Such as splitting a non-partitionable dimension of a norm operator without proper aggregation or intentional approximation; or applying misaligned masking and slicing when extracting partitioned tensor shards.
4. **Incorrect scaling of distributed states.** Such as incorrect loss scaling and gradient norm calculation—due to miscounting tensor replicas under interleaved parallelism, *e.g.*, expert and context parallelism.
5. **Incorrect pipeline scheduling.** Such as gradient synchronization occurring before the final backward iteration of a batch; or microbatches being mistakenly shuffled.
6. **Incorrect local buffer updates.** Such as updating gradient buffers only for a subset of parameters in the final step; only updating BF16 buffers during mixed precision training.
7. **Incorrect buffer management.** Violations of memory layout constraints such as contiguity, alignment, or padding required by collective communication routines.

Bug categories (1–4) can be completely eliminated by TRAINVERIFY. These communication and computation errors manifest as incorrect arithmetic logic, which is explicitly represented in the parallelized DFG. TRAINVERIFY's graphs also include loss scaling and gradient norm calculations that adhere to the original manual implementation. Thus, the violations will be detected through equivalence checking in the affected stages.

For pipeline scheduling (5), TRAINVERIFY ensures that the execution sequence is valid. Mistakenly shuffled operators that distort the intended data flow are eliminated through early analysis of data dependencies. Omitting microbatches prior to gradient synchronization is detected through the verification of finalized gradients. While TRAINVERIFY guarantees that microbatch scheduling is functionally correct, it does not enforce adherence to specific scheduling strategies—such as interleaved-1F1B—that aim to optimize pipeline efficiency.

Bugs related to buffers (6–7) are out of scope for TRAINVERIFY as discussed in § 4.

**Reproduction and Verification.** To demonstrate the effectiveness of TRAINVERIFY, we apply it to verify incorrect execution plans. We select 14 non-trivial *real-world* cases from Table 1, covering the broad categories in §8.3, as shown in Figure 9. As these cases span multiple systems, we identify the underlying causes and reproduce them via careful mutation, either in nnScaler's system code or directly in the generated plans, ensuring successful model instantiation. Evaluation is conducted on the LLaMA3-8B model using 2-way DP, TP, and PP.

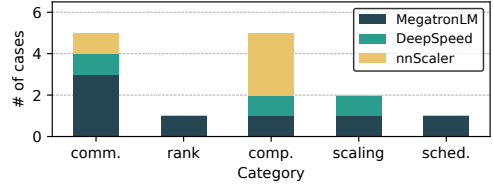

**Figure 9.** Reproduced incorrect parallelization cases.

TRAINVERIFY detects equivalence violations in all 14 plans, each completing within one minute. Two mutations, one in scheduling and one in communication, are caught early via topological sorting due to broken tensor dependencies. Two other communication mutations are flagged during reduced shape inference for invoking incorrect primitives that cause output shape mismatches, *e.g.* replacing `AllReduce` with `AllGather`. The remaining are detected during stage-parallel verification of their respective stages. Upon detection, TRAINVERIFY generates a counterexample with concrete input/output values and violated lineage for the stage.

Our Appendix B.1 includes a detailed case study for a subtle distributed computation bug in MegatronLM and how TRAINVERIFY eliminates such bugs.

**Exposing New Violations.** In applying TRAINVERIFY, we uncover two classes of buggy plans in nnScaler:

- `C1`: Sharding a non-partitionable dimension.
- `C2`: Dangling tensors in backward pass.

They are reported to and confirmed by the nnScaler developers. `C1` are discovered in its example code, where the annotation for the modularized operator `apply_rotary_embedding` is inaccurate. Under some strategies, the sequence dimension of the activations is partitioned, but the precomputed positional embeddings are not sliced accordingly, leading to sharded subsequences being misencoded as independent starts. `C2` include backward tensors with consumers but no producers, which break the data flow. Although they happen to be not reflected in the generated model code due to the autograd accumulation mechanism, it reveals potential errors in the corresponding system implementation.

## 9 Discussions

TRAINVERIFY opts to verify full execution plans without assuming homogeneity across layers or parallelism degrees. This preserves generality, accommodating to increasingly flexible parallelization strategies, *e.g.*, differentiated parallelism applied across Transformer layers [48], or across data pipelines [39].

One limitation of TrainVerify is its reliance on graph-based execution plans and tensor lineage for scalable verification, so it does not directly apply to code-based parallelization such as MegatronLM [61]. It is possible to support these systems by leveraging tools like PyTorchFX [65] to trace graphs, handling traced communication operations, and obtaining lineage information through either manual or automated annotation.

TrainVerify currently requires manual specification of shape reduction rules and operator rewrites to support symbolic tensors, though their efforts are moderate. Automating them, *e.g.*, with `autograd` and program synthesis, is a future work.

## 10 Related Work

**Neural network equivalence.** A body of work [27, 37, 40, 67] explores neural network equivalence. The closest to TrainVerify is the work by Eleftheriadis et al. [37], which encodes neural networks into SMT clauses, similar to TrainVerify, but focuses on checking approximate equivalence for knowledge distillation. Their system is limited to one- or two-layer feed-forward neural networks with a maximum layer width of 2,000 neurons. In contrast, TrainVerify scales to state-of-the-art large language models with billions of parameters.

TASO [40] is a neural network optimization system that accelerates networks by substituting certain components with more efficient alternatives. To ensure correctness, TASO includes a substitution verifier that encodes substitution rules as SMT clauses to verify the equivalence of the network before and after substitution. TrainVerify shares a similar philosophy—treating the original network as the ground truth—but differs significantly in scale. While TASO handles a small number of operators for each substitution, TrainVerify operates end-to-end, requiring verification of the entire model.

Other works address neural network equivalence for various purposes, including testing DL compiler [50], supporting model rewriting rules [27, 70], repairing models [54], and ensuring the approximation of pruned and distilled models [67]. In contrast, TrainVerify verifies equivalence and focuses on large-scale parallel training, scaling to models with billions of parameters, a level unmatched by prior work.

**Neural network verification.** In the intersection of deep learning and formal methods, neural network verification [43, 47, 71] addresses a related but orthogonal problem: verifying whether a trained neural network meets specified requirements. These systems encode the network's concrete weights and evaluate whether specific input-output pairs, potentially infinite, satisfy predefined specifications. In contrast, TrainVerify operates on symbolic tensors, verifying equivalence across all possible inputs rather than specific cases.

## 11 Conclusion

We present TrainVerify, a system that provides strong correctness guarantees for distributed training by verifying parallelization equivalence. Through multiple techniques such as symbolic representation, shape reduction, and staged verification, TrainVerify successfully scales to state-of-the-art LLMs with hundreds of billions of parameters. Our work demonstrates that formal methods can apply to and effectively benefit complex parallel training workflows in practice.

## A Implementation Details

Table 4 lists the operators supported by TRAINVERIFY. For communication operators, fused versions of the listed primitives are also adapted.

| Op Type | Rewritten Operators |
|---|---|
| **PyTorch op** | linear, sum, add, transpose, dropout, matmul, layer_norm, float, to, pow, mean, rsqrt, mul, view, div, softmax, silu, embedding, einsum, contiguous |
| **Comm. op** | chunk, move, all_reduce, all_gather, reduce_scatter, all_to_all, broadcast |
| **nnScaler op** | identity, multiref |
| **Custom op** | apply_mask, moe_gmm, apply_rotary_emb, moe_gate, create_mask, grad_norm |

Table 4. Operators supported on symbolic tensors by TRAINVERIFY

We briefly describe additional handling in shape reduction and adapting non-SIMD operators.

- For shape reduction solving, TRAINVERIFY minimizes the $L_1$ norm across input dimensions rather than the total volume of input tensors, converting the quadratic optimization goal to be linear, which alleviates solver overhead while yielding near-minimal shapes that still preserve correctness.
- For reduction rules, the mask operator requires its sequence length (i.e., the side of the mask matrix) to be $\geq 2$ to meaningfully capture triangular causal behavior. The embedding operator compresses the number of dictionary entries to be reduced $gbs \times seqlen$, matching the reduced total number of input tokens across all GPUs. The experts dimension in MoE models is not reduced.
- For adapting operators, symbolic variables pose a major challenge when used as indices in table-lookup operations. To handle the embedding layer, TRAINVERIFY assigns concrete but distinct indices to each token in the initial input sequence, enabling the lookup. The original freedom of token identity is preserved in the symbolic word vectors, so equivalence is not compromised. For the topK operator, directly replaying its logic on symbolic values is infeasible. To check its output equivalence, TRAINVERIFY instead rule-checks the participating input symbols. Then, TRAINVERIFY rewrites topK to support downstream consumption, by producing a selection matrix in the form of one-hot encodings over all experts, preserving expert selection semantics while avoiding table-lookup.

## B Evaluation Details

We include evaluation details that are omitted in the main paper. Table 5 and 6 lists configurations for models evaluated in § 8.1. Table 7 lists the configurations for models evaluated in § 8.2, where * refer to values in Figure 7.

| Parameter | 8B | 70B | 405B |
|---|---|---|---|
| layers | 32 | 80 | 126 |
| dim | 4096 | 8192 | 16384 |
| attention heads | 32 | 64 | 128 |
| sequence length | 8192 | 8192 | 8192 |
| vocabulary size | 128000 | 128000 | 128000 |

Table 5. Llama 3 Model Specifications

| Parameter | 16B | 236B | 671B |
|---|---|---|---|
| layers | 27 | 60 | 61 |
| dim | 2048 | 5120 | 7168 |
| attention heads | 16 | 128 | 128 |
| sequence length | 16384 | 16384 | 16384 |
| vocabulary size | 102400 | 102400 | 129280 |
| routed experts | 64 | 160 | 256 |
| activated experts | 6 | 6 | 8 |

Table 6. DeepSeek V3 Model Specifications

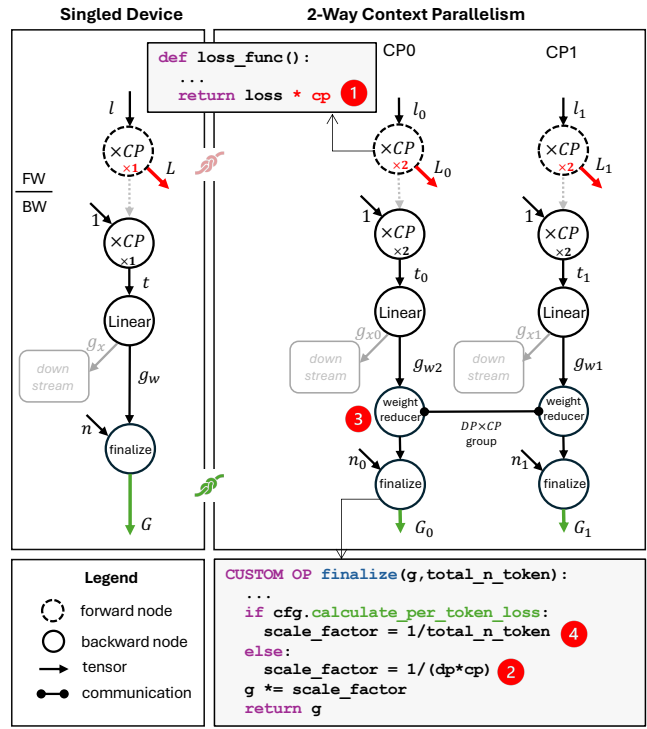

Figure 10. Gradient finalization under 2-way context parallelism. For simplicity, the producer flow for total_n_token is omitted.

### B.1 Case Study

We highlight a representative case derived from the Megatron-673 issue [18] to illustrate the subtlety of parallelization bugs and how TRAINVERIFY eliminates this broad class of silent errors. The issue leads to over-scaled distributed gradients under certain configurations. The intertwined tensor aggregation across data and context parallelism involves multiple allreduce operations over different parallel groups, with corresponding scaling and anti-scaling based on group size, token count, and the number of microbatches. The bug is triggered when `CP > 1` and `calculate-per-token-loss` is enabled, activating an execution path that mistakenly applies extra scaling—resulting in inflated loss and correspondingly over-scaled gradients.

| Group | Controlled Variable | gbs | hidden | seqlen | heads | pp | layers | dp | mb | tp |
|---|---|---|---|---|---|---|---|---|---|---|
| b | global batch size | * | 4096 | 128 | 32 | 2 | 32 | 2 | 2 | 2 |
| c | hidden dimension size | 512 | * | 128 | 32 | 2 | 32 | 2 | 2 | 2 |
| d | sequence length | 32 | 128 | * | 32 | 2 | 32 | 2 | 2 | 2 |
| e | heads | 512 | 4096 | 128 | * | 2 | 32 | 2 | 2 | 2 |
| f | pipeline parallelism | 512 | 4096 | 128 | 32 | * | 32 | 2 | 2 | 2 |
| g | layers | 512 | 4096 | 128 | 32 | 4 | * | 16 | 8 | 8 |
| h | data parallelism | 512 | 4096 | 128 | 32 | 4 | 32 | * | 8 | 8 |
| i | micro batches | 512 | 4096 | 128 | 32 | 4 | 32 | 16 | * | 8 |
| j | tensor parallelism | 512 | 4096 | 128 | 32 | 4 | 32 | 16 | 8 | * |

Table 7. Configurations for variable controlled experiments.

TrainVerify captures the parallelization logic, illustrated in Figure 10. Normally, the model scales the forward loss by `CP` ❶ to partially counteract the anti-scaling ❷ introduced by weight reducers in data parallelism ❸, ensuring that the final updates reflect per-input-sequence gradients. However, when `calculate-per-token-loss` is enabled, the averaging across the DP×CP communication group is skipped and instead replaced by averaging over the total number of trained tokens ❹. In this case, the combination of ❶ and ❹ results in the final gradients being over-scaled by a factor of `CP`.

TrainVerify eliminates such bugs by comparing data flow of shape-reduced symbolic tensors. While the violation could be detected earlier via $L == L_0$, practical implementations typically do not enforce strict equivalence on distributed losses. Moreover, adapted graphs from manually-parallelized models lack the backward lineage, e.g. $t \stackrel{\mathcal{L}}{==} (t_0, t_1)$ that is naturally preserved by auto-parallel systems. As a result, TrainVerify detects the problem as soon as it visits a weight tensor in backward pass, (e.g., $g_{w0}$) by checking that its finalized gradient $G_0$ is consistent with $G$.

Such computational issues are subtle, making diagnosis particularly challenging, especially when the code spans multiple modules. The issue post reflects a 10-day effort involving users, developers, and volunteers to reproduce the problem and identify its root cause, amid early misdiagnoses and user concerns. In the year prior to that fix, over 5 issues were filed on the same code snippets, across various training configurations; some were misreported, while others were resolved after extensive discussion. TrainVerify can effectively alleviate such challenges and ensure verified correctness.

# C  Shape reduction correctness proof

When a DNN model involves large size tensors, such as popular LLMs, it becomes infeasible for current solvers (e.g., Z3) to verify its parallel execution considering the complexity. In response, we propose a verification for the same model architecture but with reduced tensor shapes, and prove that: the verification conclusion on the shape-reduced model also applies to the original model with larger tensor shapes.

## C.1  Formulation

A DNN model consists of multiple operators, such as MatMul and ReLU, which essentially are functions with data tensors as input and output. Given input tensor(s), including tensors representing weights, activation and optimizer state, such a DNN function can produce corresponding output tensor(s). Given a DNN function $f$ that executes on a single device, there is an alternative function $g$ ($g$ is different from $f$) that can execute on either single device sequentially or multiple devices concurrently. Our goal is to verify that regardless of the inputs, $f$ and $g$ can always produce the same results—an equivalence.

**Definition C.1** (Tensor). A *tensor* is an object that generalizes scalars, vectors, and matrices to higher dimensions. Formally, an order-$n$ tensor (also called an $n$-th order tensor) is an element of the tensor space:

$$\mathcal{T} \in \mathbb{R}^{d_1 \times d_2 \times \cdots \times d_n} \quad (1)$$

where $\mathbb{R}$ represents real number field and $d_1, d_2, \ldots, d_n$ denote the dimensions along each mode of the tensor. A tensor can be represented as a multidimensional array, with components indexed as:

$$\mathcal{T}_{i_1, i_2, \ldots, i_n}, \quad i_k \in \{0, \ldots, d_k - 1\}, \quad \forall k \in \{1, 2, \ldots, n\}. \quad (2)$$

Tensors are used as primitive data in machine learning.

A tensor can be partitioned into multiple sub-tensors and may stored across multiple computational devices in a distributed computing environment. Such a distributed tensor is defined by a partitioning scheme that divides $\mathcal{T}$ into smaller sub-tensors:

$$\mathcal{T} = \bigcup_{p \in \mathcal{P}} \mathcal{T}^{(p)}, \quad (3)$$

where $\mathcal{P}$ denotes the set of computational devices, and each sub-tensor $\mathcal{T}^{(p)}$ is assigned to a specific device $p \in \mathcal{P}$. The partitioning can be done in various ways, such as:

- *Dimensional-wise partitioning*: The tensor is divided into overlapping or non-overlapping blocks along one or more modes.
- *Value-wise partitioning*: Individual elements are with partial value distributed across devices.

Distributed tensors enable storing large data and effective parallel processing for large-scale machine learning.

**Lineage information of distributed tensors.** Each partition $\mathcal{T}^{(p)}$ carries *lineage information* $\mathcal{L}(\mathcal{T}^{(p)})$, which encodes the metadata necessary for reconstructing the full tensor. This lineage information includes:

- Partitioning scheme and indexing metadata,
- Partial value occupancy.

The full tensor can be reconstructed from its distributed representation using the *merge* operation:

$$\mathcal{T} = \text{Merge}(\mathcal{T}^{(p)}, \mathcal{L}). \quad (4)$$

The *Merge* function utilizes lineage information $\mathcal{L}$ to correctly assemble the distributed tensor back into its full representation.

**Definition C.2** (Functions). A general function that operates on multiple tensors can be defined as:

$$f : (\mathbb{R}^{d_1 \times d_2 \times \cdots \times d_n}, \mathbb{R}^{d'_1 \times d'_2 \times \cdots \times d'_m}, \ldots) \to \mathbb{R}^{d''_1 \times d''_2 \times \cdots \times d''_k} \quad (5)$$

where $f$ takes one or more tensors as input and outputs a tensor of a potentially different shape.

For simplicity, we assume a single-tensor input for function $f$ throughout this proof. The proof can be naturally extended to accommodate multiple input and output tensors.

People have long observed that deep learning operators like matrix multiplication and convolution are SIMD (Single-Instruction Multiple-Data): the operation consists of repeated, homogeneous computations (the "kernel") over array elements. This SIMD characteristic is the core enabler for our shape reduction mechanism. Below, we formally define what is a SIMD function.

**Definition C.3** (Kernel function). Let $f(\mathbf{x}) \to \mathbf{y}$ be a function. A kernel function $\theta$ associated with $f$ takes a subtensor from $\mathbf{x}$ and outputs a scalar value. Formally,

$$\theta_f : \mathbb{R}^k \to \mathbb{R},$$

where $k$ is the size of the subtensor of $\mathbf{x}$.

Next, we define which input subtensor is associated with each output element.

**Definition C.4** (Dependency mapping). Consider a function $f(\mathbf{x}) \to \mathbf{y}$. A dependency mapping $\tau$ associated with $f$ is a function that maps each index $i$ in $\mathbf{y}$ to a list of indices in $\mathbf{x}$. Formally,

$$\tau_f : idx(\mathbf{y}) \to [idx(\mathbf{x})],$$

where $idx(\cdot)$ is the indexing function of the tensor.

With dependency mapping and kernel function, we define SIMD functions.

**Definition C.5** (SIMD function). A function $f(\mathbf{x}) \to \mathbf{y}$ is a SIMD function if, for each $\mathbf{y}[i]$,

$$\mathbf{y}[i] = \theta(\mathbf{x}_1, \mathbf{x}_2, \ldots, \mathbf{x}_k),$$

where $\theta$ is the kernel function of $f$, and

$$\mathbf{x}_j = \mathbf{x}[\tau(i)[j]], \quad 1 \leq j \leq k$$

where $\tau$ is the dependency mapping of $f$.

By fixing a kernel function $\theta$ and a dependency mapping $\tau$, one can define an SIMD function. We denote an SIMD function $f$ using its kernel function and dependency mapping as: $\mathbf{y}[i] = \theta_f(\mathbf{x}[\tau_f(i)])$.

## C.2 Observations: SIMD characteristics in LLM operations

Deep Neural Network (DNN) computations are characterized by their application to high-dimensional data tensors. A closer examination of commonly used DNN operations reveals that a large number of elements in the output tensor share the same computational logic, differing only in the specific input elements they process. This computational pattern aligns closely with Single-Instruction-Multiple-Data (SIMD), a concept from computer architecture.

### C.2.1 Observation 1: LLM operators are SIMD functions.
We observe that all layers in the transformer architecture are SIMD, including Feed Forward layers, Multi-Head Attention layers (without masking), Add & Norm layers, ReLU, Softmax, and Residual addition.

As an example, matrix multiplication (i.e., MatMul) can be expressed as an SIMD function. Given two matrices $A \in \mathbb{R}^{m \times p}$ and $B \in \mathbb{R}^{p \times n}$, the resulting matrix $C \in \mathbb{R}^{m \times n}$ has elements $c_{i,j}$ (short for $C[i][j]$) computed by: $c_{i,j} = \sum_{k=1}^{p} a_{i,k} \cdot b_{k,j}$. MatMul is a SIMD function because it has

- a kernel function:
  $\theta(a_{i,1}, \ldots, a_{i,p}, b_{1,j}, \ldots, b_{p,j}) = \sum_{k=1}^{p} a_{i,k} \cdot b_{k,j}$;
- a dependency mapping for each input tensor:
  $$\tau^A(i, j) = [(i, k) | 1 \leq k \leq p],$$
  $$\tau^B(i, j) = [(k, j) | 1 \leq k \leq p],$$
  where $\tau^A$ and $\tau^B$ are dependency mappings for input matrix A and B;
- and MatMul can be expressed as:
  $$c_{i,j} = \theta(A[\tau^A(i, j)] \oplus B[\tau^B(i, j)]),$$
  where $\oplus$ represents list concatenation.

### C.2.2 Observation 2: LLM operators have dependency mappings that can be expressed as linear combinations.
This property is intuitive, as the "striding" of kernel functions across tensors typically occurs at regular, constant intervals. Consequently, when the input to the dependency mapping—corresponding to the output tensor's index—changes, the resulting input indices change linearly. That is, the mapping takes the linear form: $\tau(i) = M \cdot i + b$.

For example, in the above MatMul case, the dependency mapping $\tau^A$ can be written as a linear form:

$$\tau^A(i, j) = M_A \begin{bmatrix} i \\ j \end{bmatrix} + b_A, \quad M_A = \begin{pmatrix} 1 & 0 \\ 1 & 0 \\ \vdots & \vdots \\ 1 & 0 \end{pmatrix}_{p \times 2}, b_A = \begin{pmatrix} 0 & 1 \\ 0 & 2 \\ \vdots & \vdots \\ 0 & k \end{pmatrix}_{p \times 2}$$

### C.2.3 SIMD Functions in LLMs.
Following Definition C.5, LLM operators can be formally expressed as SIMD functions. We elaborate on several as examples below.

***ReLU Redefined as SIMD Function*** The Rectified Linear Unit (ReLU) activation function can be defined in SIMD terms as follows:

Given the ReLU function:

$$\text{ReLU}(x) = \max(0, x)$$

The SIMD function $\theta$ for ReLU is:

$$\theta(x_i) = \max(0, x_i)$$

For ReLU, the input mapping function $\tau$ is:

$$\tau(i) = \{i\}$$

Thus, for each output element $y_i$ in the output tensor $\mathbf{y}$, the corresponding input element is $x_i$ from the input tensor $\mathbf{x}$.

***MatMul Redefined as SIMD Function*** Matrix multiplication (MatMul) can be defined in SIMD terms as follows:

Given two matrices $A$ of size $m \times p$ and $B$ of size $p \times n$, the resulting matrix $C$ of size $m \times n$ has elements $c_{ij}$ computed by:

$$c_{ij} = \sum_{k=1}^{p} a_{ik} \cdot b_{kj}$$

The SIMD function $\theta$ for MatMul is:

$$\theta(a_{i1}, b_{1j}, a_{i2}, b_{2j}, \ldots, a_{ip}, b_{pj}) = \sum_{k=1}^{p} a_{ik} \cdot b_{kj}$$

Here, the number of input elements $k$ for each computation of $c_{ij}$ is $2p$.

The input mapping function $\tau$ for an element $c_{ij}$ is:

$$\tau(i, j) = \{\{(i, 1), (i, 2), \cdots, (i, p)\},$$
$$\{(1, i), (2, i), \cdots, (p, i)\}\}$$
$$= \{\{(i, k) | 1 \leq k \leq p\},$$
$$\{(k, i) | 1 \leq k \leq p\}\}$$

The mapping specifies the indices in matrices $A$ and $B$ that contribute to the output element $c_{ij}$ of matrix $C$.

***Convolution as SIMD Function*** A standard 2D convolution operation can be expressed in SIMD terms as follows:

Given an input tensor $\mathbf{x} \in \mathbb{R}^{H \times W \times C_{in}}$ and a kernel tensor $\mathbf{w} \in \mathbb{R}^{K_H \times K_W \times C_{in} \times C_{out}}$, the output tensor $\mathbf{y} \in \mathbb{R}^{H' \times W' \times C_{out}}$ is computed as:

$$y_{i,j,c} = \sum_{m=1}^{K_H} \sum_{n=1}^{K_W} \sum_{c'=1}^{C_{in}} x_{i+m,j+n,c'} \cdot w_{m,n,c',c}$$

The SIMD function $\theta$ for convolution is:

$$\theta(\{x_{i+m,j+n,c'} \mid 1 \le m \le K_H, 1 \le n \le K_W, 1 \le c' \le C_{in}\})$$
$$= \sum_{m=1}^{K_H} \sum_{n=1}^{K_W} \sum_{c'=1}^{C_{in}} x_{i+m,j+n,c'} \cdot w_{m,n,c',c}$$

The input mapping function $\tau$ for convolution is:

$$\tau(i,j,c) = \{(i+m, j+n, c') \mid$$
$$1 \le m \le K_H, 1 \le n \le K_W, 1 \le c' \le C_{in}\}$$

This mapping function defines how each output element $y_{i,j,c}$ is derived from a specific subset of input elements.

***Pooling as SIMD Function*** Consider a max-pooling operation with kernel size $K_H \times K_W$ and stride $s$, applied to an input tensor $\mathbf{x} \in \mathbb{R}^{H \times W \times C}$. The output tensor $\mathbf{y} \in \mathbb{R}^{H' \times W' \times C}$ is computed as:

$$y_{i,j,c} = \max_{1 \le m \le K_H, 1 \le n \le K_W} x_{si+m, sj+n, c}$$

The SIMD function $\theta$ for max pooling is:

$$\theta(\{x_{si+m,sj+n,c} \mid 1 \le m \le K_H, 1 \le n \le K_W\})$$
$$= \max_{1 \le m \le K_H, 1 \le n \le K_W} x_{si+m, sj+n, c}$$

The input mapping function $\tau$ for pooling is:

$$\tau(i,j,c) = \{(si+m, sj+n, c) \mid 1 \le m \le K_H, 1 \le n \le K_W\}$$

This formalism captures how each pooling operation aggregates information from a specific region of the input tensor.

#### C.2.4 Dependency mapping $\tau$ in DNN
For a general tensor operation defined by a function $f : \mathbb{R}^n \to \mathbb{R}^m$, the input mapping function $\tau : \mathbb{N}^m \to (\mathbb{N}^n)^k$ defines the dependency of each output element on a subset of input elements.

In its general form, $\tau$ can be expressed as:

$$\tau(i) = \{\phi(i, j) \mid j = 1, 2, \ldots, k\}$$

where:
- $i \in \mathbb{N}^m$ is the index of the output tensor,
- $j$ indexes the $k$ input elements required to compute $y_i$,
- $\phi : \mathbb{N}^m \times \mathbb{N}^+ \to \mathbb{N}^n$ is an index transformation function that determines which input elements contribute to $y_i$.

The function $\phi$ follows specific constraints to ensure a structured and predictable mapping between input and output indices. Specifically, in DNN, $\phi$ return an ordered list of multi-dimensional output indices, preserving the computational consistency required for SIMD processing. We observe in DNN that each output index is expressible as a linear polynomial, where the input index $i$ serves as the variable, while the iterator number $j$ and other shape-related parameters—such as the dimensions of the input tensor, the convolution kernel size, and the stride—act as constant coefficients or offsets.

A key property of the linear polynomial expression of $\phi$ is that nested functions naturally preserve and combine their respective input mappings $\tau$ while maintaining the linear polynomial structure.

Thus, $\tau$ provides a unified way to describe how each output element maps to its respective input elements across various DNN operations.

### C.3 Correctness proof for shape reduction

This section establishes the correctness of TRAINVERIFY's shape reduction by proving the equivalence between two data flow graphs (DFGs) on a subset of dimensions implies equivalence across all dimensions. We denote the original and transformed DFGs—before and after applying parallelization techniques—as functions $f$ and $g$, respectively.

#### C.3.1 Prerequisite relations.
Before presenting the main theorem, we begin with two equivalent definitions that serve as the foundation for the proof.

**Definition C.6** (Mapping permutation equivalence). For two dependency mappings $\tau_1$ and $\tau_2$, we call them mapping permutation equivalence, denoted $\tau_1 \cong_P \tau_2$, if there exists a permutation function $P$, such that

$$\forall i, \quad \tau_1(i) = P(\tau_2(i))$$

Mapping permutation equivalence captures LLM operators with commutative properties, where permuting the inputs does not affect the output. Similarly, we need to define a corresponding equivalence relation for kernel functions.

**Definition C.7** (Kernel permutation-set equivalence). For two kernel functions $\theta_1$ and $\theta_2$, we call them kernel permutation-set equivalence, denoted $\theta_1 \cong_Q \theta_2$, if there exists a non-empty set $Q$ of permutation functions, such that

$$\forall P \in Q, \forall \mathbf{x}, \quad \theta_1(\mathbf{x}) = \theta_2(P(\mathbf{x}))$$

**Definition C.8** (Well-formed kernel function). We call a kernel function $\theta$ well-formed if,

$$\exists \mathbf{x}, \mathbf{x}', \forall i, \mathbf{x}[i] \ne \mathbf{x}'[i] \text{ and } \forall j \ne i, \mathbf{x}[j] = \mathbf{x}'[j]$$
$$\theta(\mathbf{x}) \ne \theta(\mathbf{x}')$$

Essentially, all elements in the inputs to the well-formed kernel functions contribute to the final output. There is no such input element that does not influenct the output.

#### C.3.2 Main proofs.
Below is the main proof for the correctness of shape reduction

**Lemma C.9.** *For SIMD functions f and g with well-formed kernel functions:*

$$\theta_f \cong_Q \theta_g \ \wedge \ \tau_f \cong_P \tau_g \ \wedge \ P \in Q \implies f = g$$

*Proof.*

$$\begin{aligned}
\forall \mathbf{x}, \forall i, f(\mathbf{x})[i] &= \theta_f(\mathbf{x}(\tau_f(i))) && \text{[by Definition C.5]} \\
&= \theta_g(P(\mathbf{x}[\tau_f(i)])) && \text{[by } \theta_f \cong_Q \theta_g \wedge P \in Q\text{]} \\
&= \theta_g(\mathbf{x}[P(\tau_f(i))]) && \text{[by tensor indexing rules]} \\
&= \theta_g(\mathbf{x}[\tau_g(i)]) && \text{[by } \tau_f \cong_P \tau_g\text{]} \\
&= g(\mathbf{x})[i]
\end{aligned}$$

Because for any input $\mathbf{x}$, $f(\mathbf{x})$ and $g(\mathbf{x})$ produce the same result, therefore $f = g$. □

Next, we prove that dimention reduction also applies to reductional opreations such as sum. A reductional function $f : \mathbb{R}^n \to \mathbb{R}$ returns a single output element from processing a reduction operation among all elements in the input tensor, with the reduction operation satisfying the commutative and associative laws.

**Definition C.10** (Reductional function). For an input tensor $X \in \mathbb{R}^n$, the reductional function $f$ applies a binary operation $\odot$ to all elements of $X$ such that:

$$f(X) = x_1 \odot x_2 \odot \cdots \odot x_n,$$

and $\odot$ satisfies commutativity ($a \odot b = b \odot a$) and associativity ($(a \odot b) \odot c = a \odot (b \odot c)$).

**Lemma C.11.** *Given reductional functions f and g,*

$$\forall \mathbf{x} \in \mathbb{R}^2, f(\mathbf{x}) = g(\mathbf{x}) \implies \forall \mathbf{x} \in \mathbb{R}^n, n > 2, f(\mathbf{x}) = g(\mathbf{x}).$$

*Proof.* The lemma can be proved using mathematical induction.

*Base case:* Consider the base case: $f(\mathbf{x}) = g(\mathbf{x}) | \mathbf{x} \in R^{[2]}$, which is true.

*Inductive Step:* Assume that $f(\mathbf{x}) = g(\mathbf{x}) | \mathbf{x} \in R^{[k]}, k >= 2$ is true. Now we need to prove that $f(\mathbf{x}) = g(\mathbf{x}) | \mathbf{x} \in R^{[k+1]}$ is also true.

- Start with $\{\mathbf{x} | \mathbf{x} \in R^{[k+1]}\}$, $\mathbf{x}$ can also be expressed as $concat(\mathbf{x}[1..k], \mathbf{x}[k+1])$.
- Given the commutative and associative laws, $f(\mathbf{x}) = f(f(\mathbf{x}[1..k]), \mathbf{x}[k+1])$, and $g(\mathbf{x}) = g(g(\mathbf{x}[1..k]), \mathbf{x}[k+1])$.
- Based on "$f(\mathbf{x}) = g(\mathbf{x}) | \mathbf{x} \in R^{[k]}, k >= 2$" is true, $f(\mathbf{x}[1..k]) = g(\mathbf{x}[1..k])$, then $g(\mathbf{x}) = g(f(\mathbf{x}[1..k]), \mathbf{x}[k+1])$.
- Based on "$f(\mathbf{x}) = g(\mathbf{x}) | \mathbf{x} \in R^{[2]}$", $g(\mathbf{x}) = g(f(\mathbf{x}[1..k]), \mathbf{x}[k+1]) = f(f(\mathbf{x}[1..k]), \mathbf{x}[k+1])$, also considering $f(\mathbf{x}) = f(f(\mathbf{x}[1..k]), \mathbf{x}[k+1])$, then $f(\mathbf{x}) = g(\mathbf{x}) | \forall \mathbf{x} \in R^{[k+1]}$ is also true.

□

With Lemma C.11, checking equivalence of operator MatMul $A \cdot B$ where input tensors $A \in R^{[m,k]}, B \in R^{[k,n]}$, with $m, k, n \in \mathbb{Z}^+$ can be simplified by setting $k = 2$.

Finally, we prove the main theorem below. Notice that TRAINVERIFY's SMT solver proves the precondition: there exists some dimension of the output tensor (namely, $i$) that both $f$ and $g$ produce the same result for any input $\mathbf{x}$.

**Theorem C.12.** *Let f and g be functions composed by LLM operators.*

$$\exists i, \forall \mathbf{x}, f(\mathbf{x})[i] = g(\mathbf{x})[i] \implies \forall i, \mathbf{x}, f(\mathbf{x})[i] = g(\mathbf{x})[i]$$

*Proof.* By the precondition, we can derive that there exists a non-empty set $Q$ such that $\theta_f \cong_Q \theta_g$. We prove this by contradiction—if $\theta_f$ and $\theta_g$ are not kernel permutation-set equivalence, then there must exist some input $\mathbf{x}'$ where $\theta_f(\mathbf{x}') \neq \theta_g(\mathbf{x}')$. Consdier this $\mathbf{x}'$, we can construct an input tensor $X$ to $f$ and $g$ such that $\mathbf{x}'$ is a sub-tensor of $X$ and $\mathbf{x}'$'s corresponding output poinsition is the ith element in the output. Because $\theta_f(\mathbf{x}') \neq \theta_g(\mathbf{x}')$, then $f(X)[i] \neq g(X)[i]$, a contradiction to the precondition.

Next, by our observation 2 and the precondition, we conclude $\tau_f \cong_P \tau_g$. From the precondition, we can derive that $\exists P, \tau_f(i) \cong_P \tau_g(i)$. By observation 2, we know that the $P$ applies to all dimensions (i.e., $\tau_f \cong_P \tau_g$) due to the linear transformation.

Then, using the precondition and $\tau_f \cong_P \tau_g$, we prove $P \in Q$ by contradiction: assume $P \notin Q$, then $\exists \mathbf{x}', \theta_f(\mathbf{x}') \neq \theta_g(P(\mathbf{x}'))$. Because the precondition is true for all input tensor, we can construct an input tensor $X$ to $f$ and $g$, such that $X[\tau_f(j)] = \mathbf{x}'$. Now we have:

$$\begin{aligned}
f(X)[j] &= \theta_f(X[\tau_f(j)]) \\
&= \theta_f(\mathbf{x}') \neq \theta_g(P(\mathbf{x}')) \\
&= \theta_g(P(X[\tau_g(j)])) = g(X)[j],
\end{aligned}$$

which is a contradiction to the precondition.

Finally, by Lemma C.9,

$$\theta_f \cong_Q \theta_g \wedge \tau_f \cong_P \tau_g \wedge P \in Q \implies \forall \mathbf{x}, f(\mathbf{x}) = g(\mathbf{x})$$

□